## EFECT: A Method to Quantify the Reproducibility of Stochastic Simulations.

T.J. Sego[1*], Matthias König[2], Luis L. Fonseca[1], Dilan Pathirana[3], Frank T. Bergmann[4], Hernán E. Grecco[5], Mauro Silberberg[5], Subhasis Ray[6], Baylor Fain[1], Adam C. Knapp[1], Krishna Tiwari[7], Henning Hermjakob[7], Herbert M. Sauro[8], James A. Glazier[9], Reinhard C. Laubenbacher[1], Rahuman S. Malik-Sheriff[1,7,10†]

[1]Department of Medicine, University of Florida, Gainesville, FL, USA

[2]Institute for Theoretical Biology, Humboldt University Berlin, Berlin, Germany

[3]Life and Medical Sciences Institute & Bonn Center for Mathematical Life Sciences, University of Bonn, Bonn 53115, Germany

[4]BioQUANT, Heidelberg University, 69120 Heidelberg, Germany

[5]Universidad de Buenos Aires, Facultad de Ciencias Exactas y Naturales, Departamento de Física and CONICET - Universidad de Buenos Aires, Instituto de Física de Buenos Aires (IFIBA). Buenos Aires 1426, Argentina

[6]Centre for High Impact Neuroscience and Translational Applications (CHINTA), TCG CREST, Kolkata, India

[7]European Molecular Biology Laboratory, European Bioinformatics Institute (EMBL-EBI), Hinxton, Cambridge, UK

[8]Department of Bioengineering, University of Washington, Seattle, WA, USA

[9]Department of Intelligent Systems Engineering and Biocomplexity Institute, Indiana University, Bloomington, IN, USA

[10]Department of Surgery & Cancer, Faculty of Medicine, Imperial College London, London, UK

* timothy.sego@medicine.ufl.edu

† sheriff@ebi.ac.uk

### Abstract

Reproducibility is a fundamental requirement for validating scientific claims in computational research. Stochastic computational models are widely used in fields such as systems biology, financial modeling and environmental sciences. However, achieving reproducibility in stochastic simulations remains challenging, as each run can produce different outcomes due to inherent randomness introduced by the simulation algorithm or stochastic sampling. Existing infrastructure and software tools support various aspects of reproducible model development, application, and dissemination, but do not adequately address independent reproduction of simulation results that underpin scientific conclusions. Without such independent reproducibility, results and conclusions lack credibility, as it remains unclear whether observed findings reflect genuine model behavior or are merely artifacts of stochastic variation or an underpowered study. To bridge this unmet gap, we introduce the Empirical Characteristic Function Equality Convergence Test (EFECT), a data-driven, statistically robust method to quantify the reproducibility of stochastic simulation results. EFECT employs empirical characteristic functions to compare reported results with those independently generated by assessing distributional inequality, termed EFECT error, a normalized metric (ranging from 0 to 2) to quantify how closely two distributions match. Additionally, we establish the EFECT convergence point, a quantitative metric for determining the required number of simulation runs to achieve an EFECT error value of *a priori* statistical significance, setting a reproducibility benchmark. EFECT is applicable to all bounded, real-valued outputs, regardless of the model type or simulation method that produced them, accommodating stochasticity from intrinsic model variability and random sampling of model inputs. We rigorously tested EFECT with over 40 different diverse use cases including stochastic differential equations, agent-based models, and Boolean networks to demonstrate its broad applicability and effectiveness. EFECT standardizes stochastic simulation reproducibility, establishing a workflow that guarantees reliable results, supporting a wide range of stakeholders, and thereby enhancing validation of stochastic simulation studies, across a model's lifecycle. To promote future standardization efforts, we are developing the open-source Stochastic Simulation Reproducibility software library (libSSR) in multiple programming languages for easy integration of EFECT.

## Introduction

Various disciplines have acknowledged a crisis of reproducibility in science (1), to which computational research is not immune (2). Reproducibility has been argued to be the minimum standard for assessing scientific claims in computational science (3). For stochastic modeling studies, reproducibility is essential to ensure reliability in critical predictions, from environmental assessments to financial forecasts. In deterministic computational research, reproducibility can be quantitatively assessed. In contrast, stochastic simulation studies lack a general framework for reproducibility evaluation, leaving the true extent of the problem unknown and likely underestimated due to the intrinsic variability of such systems and the absence of standardized assessment methods.

In life sciences, significant efforts are made through the development of an ecosystem of community infrastructures and standards to improve model credibility by ensuring simulation tools yield consistent outputs for identical inputs (4). These efforts include resources like BioModels, the world's largest repository of curated biological models (5), and BioSimulators, a free online registry of biological model simulators (6), both of which leverage community standards such as the Systems Biology Ontology (SBO, (7)), Systems Biology Markup Language (SBML, (8)), and Simulation Experiment Description Markup Language (SED-ML, (9)), to promote reusability and reproducibility of biological models in diverse fields including neuroscience (10), quantitative systems pharmacology (11), and drug discovery (12). A 2021 study showed that, despite these efforts, nearly half of 455 analyzed deterministic ordinary differential equation models from peer-reviewed articles cannot be reproduced (13). Given that stochastic simulation yields varying results with every run, it was expected that their reproducibility would be worse.

Best practices for reproducible computational research emphasize that researchers must, at minimum, reproduce their own results (14). In systems biology, guidelines recommend storing all raw simulation data, sharing data underlying graphs and tables, and automating model verification and validation (15), with some advocating for journal policies mandating raw data publication (16). Additional recommendations include ensuring statistical repeatability of simulation results and expanding workflows to validate reproducibility (17). Most guidelines focus on deterministic models, yet advanced research in systems biology, finance, and environmental sciences increasingly uses stochastic models to capture random processes that deterministic approaches cannot (18–20).

Emerging technologies like digital twins, which will likely use stochastic modeling, need rigorous verification and validation of simulation results (21). There is a growing emphasis on analysis of sample distributions of simulation results to ensure model credibility (22). Reproducibility in stochastic modeling is crucial for reliable environmental predictions, such as climate models, where variability and uncertainty must be systematically addressed (23). Here, reproducibility refers to obtaining consistent results from independent studies under similar experimental conditions (16,24). Stochastic simulation results, influenced by intrinsic variability or input sampling (e.g., parameters, initial conditions), vary with each run. This inherent variability of results necessitates robust metrics and standards to quantify reproducibility and guarantee that outcomes will not be different when a study is repeated. This is essential to ensure findings are reliable and consistently validated across studies, especially when findings influence decisions pertaining to large social and economic impact.

Reproducing individual stochastic simulation trajectories of a study is feasible using the seed from pseudorandom number generators, though there are challenges due to software implementation differences (17). However, this approach has two major limitations: (i) it overlooks underpowered or selective model simulation sampling in the study, weakening assessment, and (ii) storing and sharing data for all individual trajectories does not scale up to community-level practice. Alternatively, comparing stochastic results via distributional statistics requires assumptions (e.g., normality) that limit applicability outside specific contexts. For instance, the SBML Test Suite (25) compares stochastic results using summary statistics. However, summary statistics like mean and standard deviation are not unique and so are unreliable for general identification of differences in stochastic model results (*e.g.*, when initial conditions have different distributions with the same summary statistics, Source Code S3.20). Previous work developed a test and heuristics to quantify aleatoric uncertainty for determining a minimum sample size that sufficiently represents a stochastic model (26,27) but did not provide a means to support standardized reproduction of results.

In this work, we developed a method and metrics spanning the lifetime of a model by which stochastic simulation results can be quantitatively reproduced. Our method tests for equality of stochastic results by evaluating whether results converge to the same unique, unknown distribution using empirical characteristic functions. Hence, we call our method the Empirical Characteristic Function Equality Convergence Test (EFECT). We showed the broad relevance of our method by demonstrating its application with multiple stochastic modeling applications, modeling methodologies, and BioModels entries. Our method quantifies differences in stochastic simulation results independent of modeling method and with universal meaning, such that outcomes from reproducing results are generally interpretable. We designed a workflow built on our method to accommodate data exchange within modeling communities for stochastic model reproducibility through outlets like online repositories and journal publications.

## Results

EFECT-based reproduction of stochastic simulation results consists of two components: first, generation of reliable and reproducible stochastic simulation results by the modeler; second, testing whether the stochastic simulation results have been sufficiently reproduced using a well-defined quantitative metric by a model curator. We refer to the individual who produces the results as the Modeler and to the one who independently analyzes the model and reproduces the results as the Curator. We require that EFECT assesses similarity of results using a quantitative metric, called the *EFECT error*, with meaning that is both independent of a model's mathematical construct and scale of results, and robust to support a broad range of models and sources of results' stochasticity. In particular, we require that EFECT does not rely on any knowledge about the model from which results were produced. Likewise, we require that the EFECT error supports results with variance produced by simulation stochasticity and/or stochastic selection of model parameters and initial conditions. EFECT assesses the similarity of stochastic simulation results by testing for equality in distribution of results through comparison of empirical characteristic functions (ECFs) generated from results distributions. Throughout, we refer to a set results collected from one or more executions of a stochastic simulation as a *simulation sample*, and likewise to the number of executions as the *sample size*. The EFECT error quantifies differences of simulation samples over a model- and scale-independent range, where an EFECT error of zero demonstrates equality of simulation samples, and of two demonstrates maximum difference. Using EFECT, a Modeler performs a *test for reproducibility* to test whether the distribution of their simulation sample is reproducible by randomly dividing their simulation sample into two subsamples of equal size and measuring the EFECT error of the subsamples. The Modeler improves the reproducibility of their simulation sample by repeating this process while increasing sample size until the distribution of the sampled EFECT error satisfies a reproducibility criterion, called the *EFECT convergence point*. We refer to the minimum simulation sample size to satisfy the EFECT convergent point as the *EFECT sample size*. Given the size of the subsamples, the ECFs generated from one of them, and some additional information (described below) accompanying a reported model, a Curator tests whether they can reproduce a simulation sample by comparing to ECFs from their own implementation of the reported model. A Curator tests a reported simulation sample to their own for equality in distribution according to the null hypothesis that samples of the EFECT error when comparing simulation samples and when quantifying reproducibility are taken from the same unknown distribution. A Curator assesses whether they reproduce a simulation sample by rejecting the null hypothesis at an *a priori* significance level according to Chebyshev's inequality for unknown population mean and variance. For details, see *Materials and methods*. For a summary for all test scenarios, see Table S1. For details on test scenarios and locating their supplementary materials, see Table S2.

*EFECT Accurately Quantifies Similarity in Stochastic Simulation Results*

To evaluate the robustness and generality of EFECT, we rigorously tested it across more than 40 diverse use cases, including stochastic differential equations, agent-based models, and Boolean networks, encompassing both intrinsic stochasticity and variability introduced by input sampling. A full list of these test scenarios is provided in Table S2. Below, we summarize these use cases and highlight EFECT's ability to quantify reproducibility across a range of modeling approaches, simulation platforms, and sources of stochasticity.

<u>Stochastic Algorithms and Sampling of Parameters and Initial Conditions</u>

We first tested the capability of EFECT to quantify the similarity of simulation samples using a variety of ordinary differential equation (ODE) models with different sources of stochasticity (Figure 1). Test models consisted of either deterministic simulation while sampling model parameters and/or initial conditions using a normal distribution for one, two, and ten parameters and/or initial conditions, or using a uniform distribution sampling for one or ten parameters and/or initial conditions; or of stochastic simulation using the Gillespie Stochastic Simulation Algorithm (28,29), including two BioModels entries (30,31). For all test models subject to a source of stochasticity, we increased sample size and quantified reproducibility of the simulation sample by evaluating the similarity of equally sized subsamples as previously described for our Modeler's test for reproducibility. Similarity of simulation samples was quantified using the EFECT error defined in [ 3 ]. We found that the EFECT error decreased with increasing sample size for all modeler's test cases according to a power law. A representative result of this general observation is shown in Figure 1D. EFECT reliably quantifies reproducibility of stochastic results from ODE simulations, with reproducibility improving systematically as sample size increases.

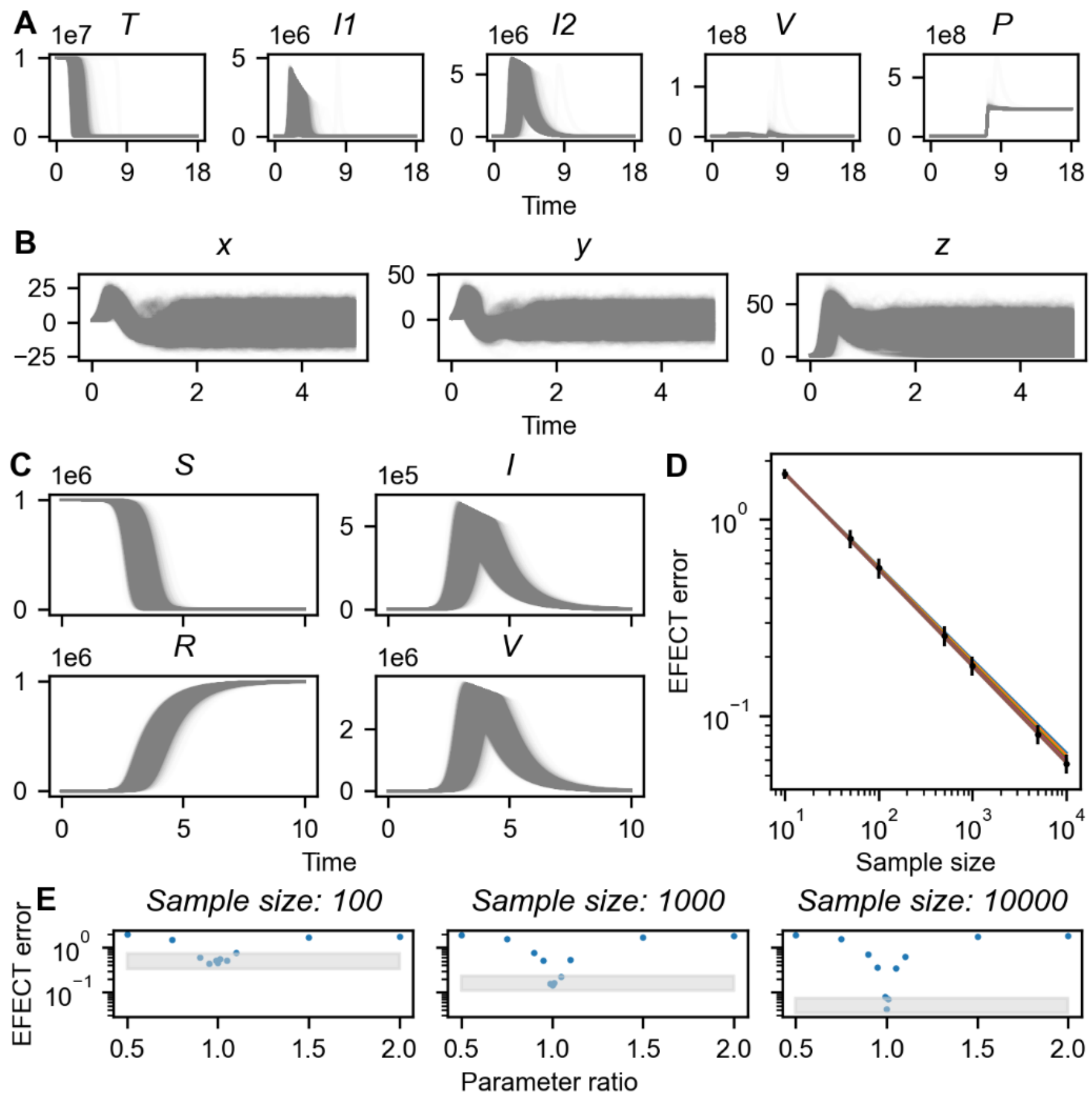

Figure 1. Testing similarity of stochastic biological simulation results from ordinary differential equation models. A: Trajectories while randomly sampling a parameter of a model of viral-bacterial coinfection with cell populations *T*, *I1*, *I2*, and viral and bacterial loads *V* and *P* (Source Code S3.1). Sample size of 10k trajectories shown. B: Trajectories while simulating a Lorenz system with the Gillespie stochastic simulation algorithm (Source Code S3.9). Sample size of 10k trajectories shown. C: Trajectories while randomly sampling a parameter of a model of viral infection. Sample size of 10k trajectories shown. D: EFECT error quantifying reproducibility of a sample from the model shown in panel C as a function of sample size. An EFECT error of zero demonstrates equality, and of two demonstrates complete dissimilarity. Lines show fits of a typical power law relationship, where similarity increases with increasing sample size. Dots show mean evaluations of the EFECT error. Bars show one standard deviation of EFECT error evaluations upward and downward. E: EFECT error per sample size from the model shown in panel C while comparing samples from models of the same structure but a difference in parameter (as measured by the parameter ratio). EFECT error ranges from 0 (identical distributions) to 2 (maximally different distributions). Shaded regions show the mean ± three standard deviations of the EFECT error when testing for reproducibility.

Detecting Parametric and Structural Differences

We then tested the capability of EFECT to detect structural and parametric differences in SDE models by comparing simulation samples generated from them. We tested detection of structural differences by comparing simulation samples of various sample sizes generated from different models. When comparing simulation samples from models with structural differences, we found that the EFECT error does not decrease with increasing sample size but instead tends to near maximum difference (Source Code S3.13). We tested detection of parametric differences by comparing simulation samples of various sizes generated from the same model but with a difference in the distribution from which a parameter value was drawn. Specifically, we assessed the likelihood of a false-positive according to [ 6 ] when testing for equality in distribution for a given sample size and difference in parameter distribution. Generally, EFECT can detect a difference if it is unlikely to yield a false-positive. In this test, we perform the following procedure: Step 1: generate a simulation sample; Step 2: quantify reproducibility of the simulation sample; Step 3: extract a subsample of half the size (as would be done when reporting a model); Step 4: generate a second simulation sample of the same half size and with a modified parameter distribution; and Step 5: test how well the simulation sample from Step 4 reproduces the simulation sample from Step 3.

For the model shown in Figure 1C, we scaled the mean and standard deviation of the normal distribution for a single parameter by factors ranging from 0.5 to 2 (Source Code S3.14). Using a significance level of 0.05, we found that EFECT could detect differences in distributions for a sample size of 10,000 when scaling downward or upward by as little as 5% to 1%, respectively. The same test and model with a sample size of 1,000 could detect 5% scaling differences. With a sample size of 100, the same test and model could detect 25% downward or 50% upward scaling differences. See Table S1 for computed p-values. The same analysis was repeated the following: with different ODE models (detected 5% or more scaling differences for sample size 10,000, Source Code S3.17); with constant parameters with a scaling difference and stochastic simulation (detected 5% or more scaling differences for sample size 10,000, Source Code S3.18); or while individually scaling distribution mean (detected 5% or more scaling differences for sample sizes 5,000 and 10,000, Source Code S3.16) or variance (detected scaling differences of 25% or more downward, and 10% or more upward, for a sample size of 10,000, Source Code S3.19). In all tests, upward and downward scaling differences of between 5% and 50% were detectable with a significance level of 0.05 when the reproducibility test EFECT error mean was below 0.1. Hence, EFECT can detect even a single parametric difference between models by comparing samples from them, and increasingly so with decreasing EFECT error.

Reproducibility with Stochastic Agent-Based and Boolean Network Models

To demonstrate whether EFECT supports reproducibility of stochastic results from methods other than SDEs, we tested whether we could perform the same analysis with three implementations of an agent-based model (ABM, Figure 2). We considered the Grass-Sheep-Wolf (GSW) ABM implemented in NetLogo ((32), Figure 2A), and implementations derived from it in MATLAB and Python (Source Code S4). We generated samples with sizes up to 20,000 (Figure 2B) and again found a decrease in EFECT error with increasing sample size according to a power law when testing for reproducibility (Figure 2C). Comparing results from different implementations, EFECT showed that no two implementations produce results that are equal in distribution (Figure 2D). Thus, we found that results from the three ABM implementations were reproducible, and that none of them produced the same results, according to EFECT. We also tested a stochastic Boolean network model of cell fate (33) implemented in MaBoSS (34) and found that EFECT supports evaluation of the reproducibility of stochastic Boolean network model results (Source Code S2.1, S3.26). EFECT distinguishes between different implementations of the same agent-based model and quantified reproducibility of the stochastic Boolean network model, demonstrating its utility for verifying reproducibility independent of modeling approach.

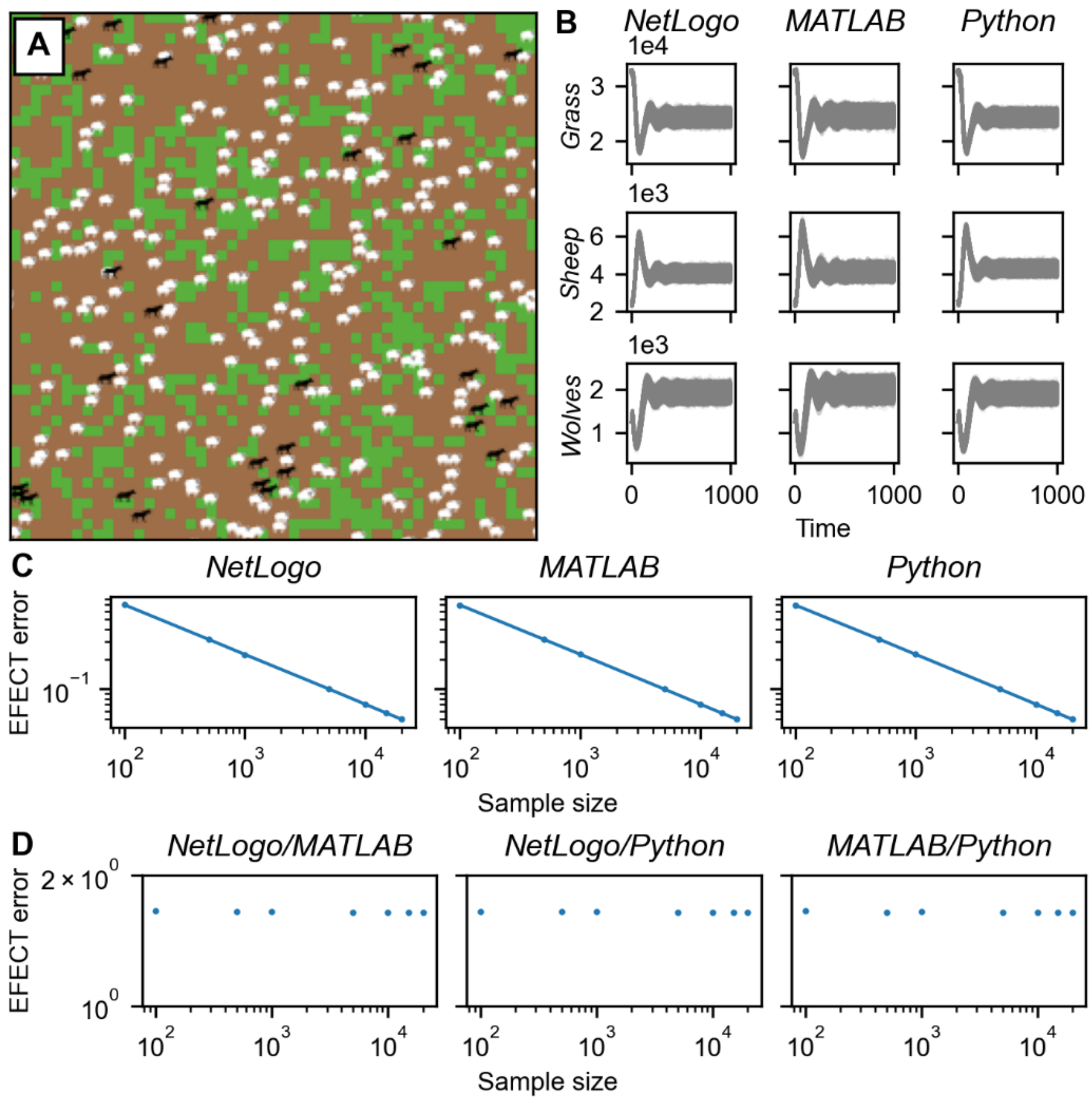

Figure 2. Testing reproducibility of results from three implementations of a Grass-Sheep-Wolf agent-based model. A: Screenshot of the original agent-based model implemented in NetLogo. B: Trajectories of model Grass (top), Sheep (middle), and Wolves (bottom) from 20,000 executions of the NetLogo (left, Source Code S4.2), MATLAB (center, Source Code S4.1), and Python (right, Source Code S4.3) implementations. C: Testing for reproducibility of results distributions from each implementation (Source Code S3.21). Results from all implementations converged in distribution. D: Testing for equality in distribution of pairs of samples from all implementations. No two implementations produced samples that were equal in distribution, as demonstrated by non-converging EFECT errors.

Robustness of EFECT to Model Complexity and Scale

Finally, further evaluation showed that testing for reproducibility with EFECT is generally insensitive to the number of model variables (Source Code S3.30, Figure S1), significant differences in scale between model variables (Source Code S3.27, Figure S2), or order of magnitude of standard deviation (Source Code S3.31). This demonstrates that EFECT is a robust method to test model reproducibility. We also found that the runtime of our implementation of EFECT increases linearly with the number of model variables (runtime of 305 ± 86 s, 401 ± 119 s, 469 ± 134 s, 642 ± 232 s, and 1,261 ± 484 s when testing reproducibility for 2, 3, 4, 5, and 10 model variables on an Intel i9, Source Code S2.2). EFECT maintains consistent performance across models with varying numbers of variables, scales, and variances, highlighting its robustness and generalizability.

*Workflows for Reproducibility of Stochastic Models*

After testing that EFECT supports quantifying whether simulation results are reproducible and reproduced, we developed a Stochastic Simulation Reproducibility (SSR) workflow that defines the steps required by all stakeholders (*i.e.*, the Modeler and Curator) over the lifetime of a model to generate stochastic results that are independently reproduced. Figure 3 graphically demonstrates these steps, the details of which are described in Materials and methods. Generally, the workflow involves the Modeler producing a published model and supporting data, which the Curator uses to test and report whether they can reproduce results according to an *a priori* significance level. Generally, neither stakeholder is required to use the same simulator that implements the model (though it may be beneficial to use different simulators).

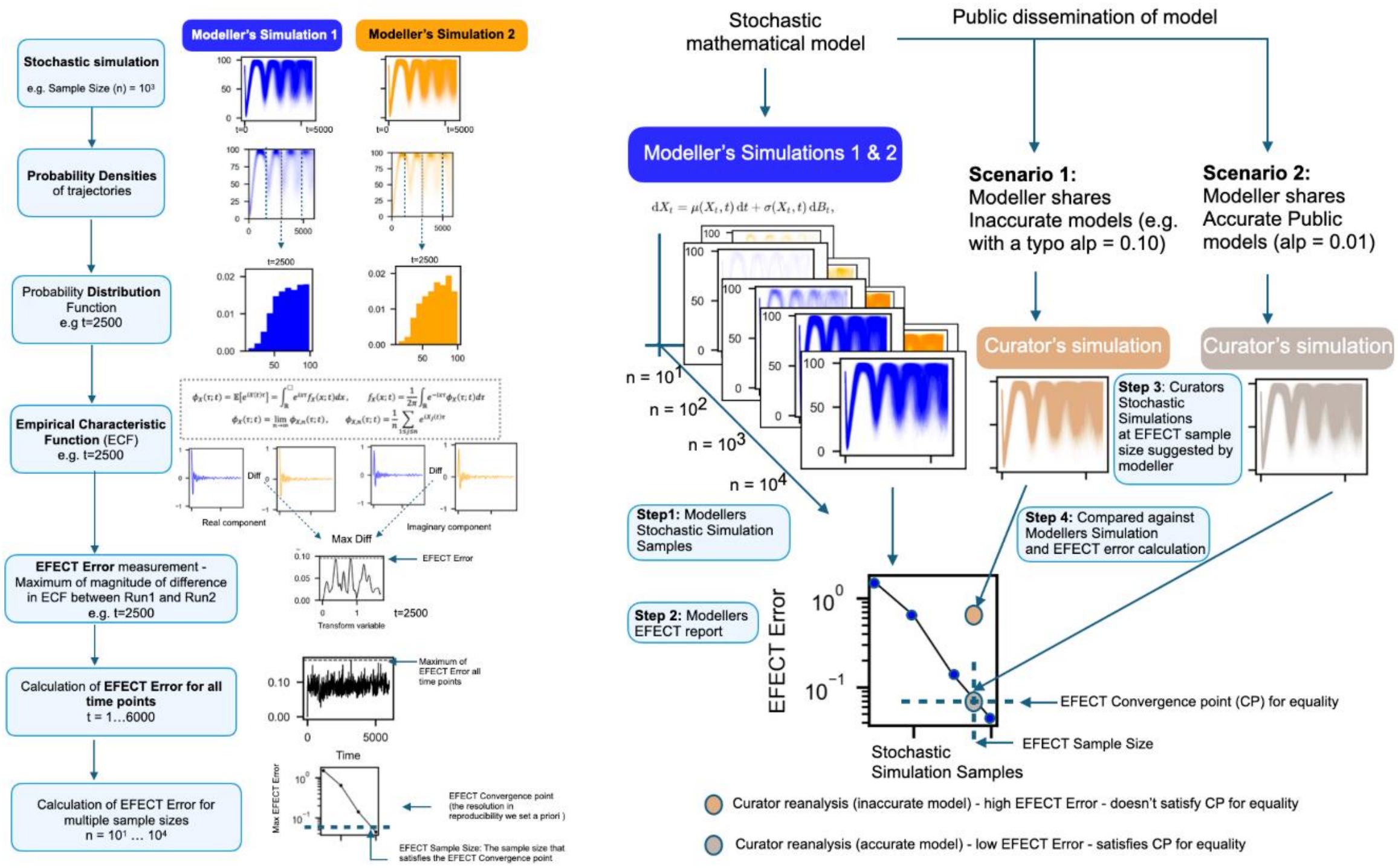

Figure 3. Graphical representation of the steps involved in stochastic simulation results comparison using EFECT. Left: Demonstration of each step in the process of modelers' internal reproducibility exercise, comparing two independent stochastic simulation results using EFECT. A modeler samples a stochastic simulation and produces empirical characteristic functions (ECFs) of them for testing whether their sample is equal in distribution to another sample by comparison of ECFs. The modeler uses the EFECT Error to compare the ECFs and quantify reproducibility of their results for each sample size. The modeller increases sample and continues the comparison until a threshold, the EFECT Convergence Point, is achieved. Right: Two possible scenarios during a reproducibility study using EFECT, where differences in results are detected (Scenario 1) and results are verified as reproduced (Scenario 2). Having achieved an established EFECT Convergence Point, results can be compared in a statistical test of whether two given results samples are realizations of the same stochastic process. Disproving the hypothesis with an *a priori* significance indicates that results are not reproduced.

A Modeler iteratively adds simulation results to a sample until results are sufficiently reproducible. The Modeler evaluates reproducibility by randomly generating two equally sized subsamples of their current sample and calculating the EFECT error when comparing ECFs of the subsamples. Evaluations of the EFECT error are collected as a random variable and the process of randomly selecting and evaluating subsamples is repeated until the absolute relative change in the mean of EFECT error evaluations converges below a threshold (here tested with a threshold of 0.1%). Reproducibility is determined according to a threshold for the statistics of the EFECT error. Here we assume that the modeler, curator, and others in their field have defined such a threshold, an *EFECT convergence point*, through consensus (here accepted when the mean plus three standard deviations is below 0.075). When a sample is not reproducible, the Modeler generates more results, adds them to the sample, and repeats the process. When a sample is reproducible, we recommend that the Modeler generates minimal supporting data, called an *EFECT report*, to allow others to reproduce a subsample. Excluding relevant information specific to a model, simulation, or data formatting other than what our workflow requires, the contents of an EFECT report is as follows (see Table S3 for examples):

- *Stochastic variable names*: A list of all named stochastic variables in reported results.
- *Simulation times*: A list of all simulation times at which results are reported.
- *EFECT error statistics*. The accepted mean and standard deviation of the EFECT error during the test for reproducibility.
- *EFECT sample size*: The size of an accepted sample when testing for reproducibility.
- *ECF evaluations*: Evaluations of ECFs generated from a subsample of an accepted sample for each named stochastic variable at each simulation time.
- *ECF domains*: Maximum transform variable value for all reported ECF evaluations according to [ 5 ].
- *Significant figures*: The significant figures of results produced by the employed simulator(s) when testing for reproducibility.

A Curator develops their own implementation of a published model and generates results distributions according to information provided by the EFECT report accompanying the model. Specifically, the sample generated by the Curator will be of a size equal to the reported sample size and record results for all named stochastic variables at all reported simulation times. The Curator performs the test for reproducibility and generates their own EFECT error statistics. Finally, the Curator randomly selects a subsample, computes the EFECT error for comparing the generated ECFs to those reported in the EFECT report, and computes the probability that the EFECT error when comparing ECFs was taken from the (unknown) distribution of the EFECT error during the Curator's test for reproducibility. The Curator determines an outcome according to whether the computed probability is above (reproduced) or below (not reproduced) a significance level. A process diagram graphically illustrates these steps in Figure S3.

To test whether our SSR workflow and EFECT report sufficiently support a lifecycle of model development that includes reproducing stochastic simulation results, we simulated the processes defined for a Modeler and Curator and exchange of information between them for an ODE model of viral infection (35) defined in SBML (Figure 4). In these simulations, a Modeler reports results from an ODE model with random parameter sampling as generated by libRoadRunner (36), and a Curator reproduces results using COPASI (37). In our simulation of the processes performed by the Modeler (*i.e.*, the "Modeler Workflow" in Figure 3), the Modeler begins by iteratively increasing the size of their sample and testing for reproducibility until a power law could be fit to EFECT error evaluations. The Modeler rounds simulation results to six significant figures for compatibility with COPASI default output (neglecting this step results in failure to reproduce, Source Code S3.25). The fit is used to estimate the required sample size to achieve an EFECT convergence point, here defined *ad hoc* as an EFECT error with mean plus three standard deviations below 0.075, when testing for reproducibility (EFECT error of $54.4\times10^{-3} \pm 6.40\times10^{-3}$, Figure 4A, B, C), the final size of which was 9,644. An EFECT report is then generated and stored to file (Source Code S3.22). In our simulation of the processes performed by the Curator (*i.e.*, the "Curator Workflow" in Figure 3), the Curator implements the same model in COPASI, generates a sample of the reported size for all reported stochastic model variables and simulation times, and performs the test for reproducibility (EFECT Error of $53.5\times10^{-3} \pm 6.59\times10^{-3}$). The Curator then compares the ECFs generated from a randomly selected subsample to those reported in the EFECT report. Our simulation showed that the Curator accepts that results are reproduced with EFECT error 0.0568 and p-value 1 (Figure 4D, Source Code S3.23) and can readily detect differences in model parameter distributions (Source Code S3.24).

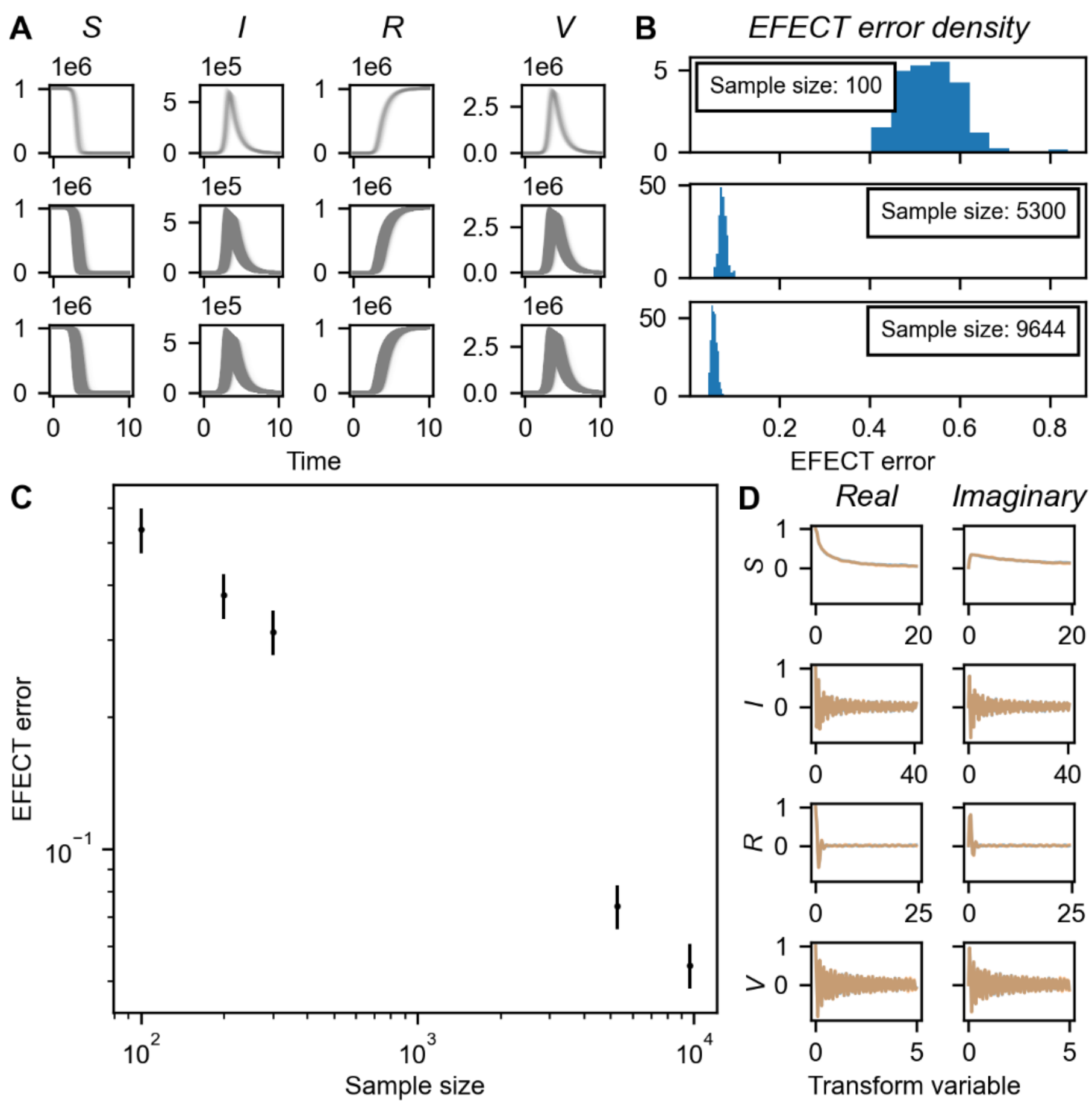

Figure 4. Simulation of the workflow for reproducing stochastic simulation results using different simulators. A, B, C: Modeler workflow using libRoadRunner. Trajectories (panel A) and corresponding EFECT error sampling (panel B) per select sample size are shown when testing for reproducibility (panel C). A sample was considered reproducible when the EFECT error plus three standard deviations was below a threshold (here 0.075). Error bars show three standard deviations. D: Comparison of empirical characteristic functions during the Curator workflow. Curator results were generated using COPASI to generate empirical characteristic functions (orange) for comparison to those reported from the Modeler workflow (blue). The two sets of empirical characteristic functions cannot be individually distinguished because they are nearly identical. Transform variable values are multiplied by a factor of $10^5$.

*Reproducible Stochastic Simulation in Select Domains of Interest*

Finally, we performed a preliminary survey of the reproducibility of stochastic simulation results in select modeling domains of broad interest (pandemic epidemiology, finance, and physics-informed neural networks). In all cases, we first attempted to qualitatively reproduce reported results, and then assessed the extent to which reported results can be reliably reproduced according to the EFECT error.

We selected six published stochastic differential equation models of the COVID-19 pandemic (38–43) and attempted to reproduce 14 reported figures (Table S2, Source Code S3.32-S.37). Of these, we could qualitatively reproduce four figures and another four figures with corrections but could not reproduce the other six. Only one figure reported results from a sample of sufficient size to produce an EFECT error on the order of our *ad hoc* EFECT convergence point of 0.075 (sample size 50,000 reported, computed EFECT error less than 0.05 for all experiments with a sample size of 10,000). For all attempted figures, we were able to generate reproducible results with EFECT error around 0.075 or less with a sample size of 5,000. One study (42) reported details on parameter fitting to data with a sample size of 20. When evaluating the reproducibility of results using the model and fitted parameters from this study, we found that sample sizes as great as 1,000 allow selecting any value for one of the fitted parameters within its bounds without producing detectable differences. We also found that a sample size of 10,000 would have made parametric differences of at least 35% detectable.

We assessed the reproducibility of results from demonstrations of financial model parameterization using Maximum Likelihood estimation with Efficient Importance Sampling (44). In summary, the approach uses adapted Markov Chain Monte Carlo (MC) sampling to estimate marginal parameter distributions for a given model and target data, a common procedure in Bayesian inference (45). We were able to qualitatively reproduce results from two Stochastic Volatility models (Source Code S3.38). We then evaluated the reproducibility of results using fitted parameters for both models, which were determined using a MC sample size of 50. With this sample size, we found that we could increase or decrease each fitted parameter by two to three standard errors without producing a detectable difference in results, whereas sample sizes of 5,000 to 10,000 would have made all parametric differences within one standard error detectable.

Last, we assessed the ability to adapt EFECT for evaluating reproducibility of partial differential equation (PDE) results during stochastic sampling when performing uncertainty quantification in physics-informed neural networks (PINNs, (46)). In summary, PINNs use deep neural networks to efficiently solve PDEs, uncertainty quantification of which requires training on results from stochastic sampling of PDE model parameters. EFECT can support such approaches for PDEs that are discretized using a fixed grid, where the PDE measurement at each grid point is treated like a model variable. We could qualitatively reproduce results in (46) and computed an EFECT error of 0.170 ± 0.0156 for the reported sample size of 1,000 used in uncertainty quantification, and of 0.0538 ± 0.00516 for a sample size of 10,000 (Source Code S3.39). We tested whether we could detect differences in four model parameters and found that the reported sample size allows decreasing one model parameter by 50%, or increasing it by 100%, without producing detectable differences in results, whereas a sample size of 10,000 allows detection of both differences.

## Discussion

EFECT is agnostic to modeling methodology, broadly supports analysis of stochastic simulation results, and is readily computable. Its methodological constraints are that evaluated results must be representable as vectors of evaluations of bounded, real-valued random variables. The interpretability of the EFECT error is independent of modeling methodology, ensuring that reproducibility metrics are interpretable and comparable across diverse model types. Our workflow produces a minimal data set supporting reproducible simulation without requiring storage of all simulation data (which would be required for comparing distributions via the Kolmogorov-Smirnov statistic and comparable), a major advantage for promoting reproducible stochastic simulation as a broadly accepted hallmark of credible stochastic modeling. While SBML and SED-ML standardize model structure and simulation procedures, EFECT uniquely addresses statistical reproducibility of simulation outcomes. Our SSR workflow with EFECT enables testing for reproduced simulation results according to a significance level, which could be determined *a priori* through consensus in various modeling communities with different expectations concerning tradeoffs in accuracy and computational cost.

We evaluated the capability of EFECT to detect differences in stochastic simulation results due to a variety of causes. Among those tested, a structural difference between models is easiest to detect, and a single parametric difference is hardest to detect. The degree to which a parametric difference can be detected in stochastic simulation results depends on the sensitivity of the assessed model outputs to the parameter. This observation is deeply connected to the structural identifiability of a model and its parameters. It naturally follows that structurally nonidentifiable models permit scenarios where different parameter sets can produce indistinguishable outputs (47). Being a data-driven approach, EFECT cannot address identifiability issues, which are better reserved for model analysis techniques (48,49). Rather, EFECT provides a statistical foundation and workflow for modeling communities to develop standards and best practices for reproducible stochastic simulation results.

SBML Test Suite has served the SBML community for verifying implementations of the Gillespie Stochastic Simulation Algorithm for nearly twenty years. EFECT can serve as the foundation for similar verification of simulators implementing other stochastic simulation algorithms. To this end, the EFECT report can be incorporated into the SED-ML standard. For example, in reference to SED-ML Level 1 Version 4, data for each model variable can be taken from DataSet entries specified in SEDBase::Output::Report::ListOfDataSets, where variable objects of corresponding data generators are scalar and output data includes simulation times. Sample size can be specified through a RepeatedTask. Parameter sampling (if any) can be specified through a RepeatedTask (*i.e.*, via SetValue) for all distributions supported by SED-ML (*i.e.*, those supported by MathML). Thus, specifying a SED-ML script for data generation during a Curator's workflow naturally follows and could be automated using SED-ML library implementations (e.g., libSEDML in C++, Java, Python, and other programming languages). It would also naturally follow that EFECT could be made broadly available as a public service for stochastic simulation verification and reproducibility in BioSimulators.

The computational cost of EFECT can be non-negligible when testing for reproducibility, depending on sample size and variance of results. With the prototype implementation in Python used in this work, tests for reproducibility in all test cases ranged over orders of seconds to hours. Furthermore, the sample sizes required to produce reproducible results may be prohibitive for computationally expensive simulations, though various modeling communities can decide what level of accuracy (*i.e.*, a significance level) is appropriate for the typical computational cost of producing simulation results. Future work should develop infrastructure supporting usage of EFECT through our SSR workflow, including a markup language, publicly distributed libraries in popular scientific programming languages (*e.g.*, C++, Python, R, Julia), hardware-accelerated algorithms, and online web services. Related strategies have been previously demonstrated for accelerating ECF computations (50). To this end, we have started such a software project, called "libSSR" (github.com/tjsego/libSSR), which is intended to provide such infrastructure in anticipation of future standardization efforts on stochastic simulation reproducibility. Current work is providing support in Python. Early work in application with libSSR has demonstrated supporting reproducibility studies (Source Code S3.40, S3.41, S3.42) with COPASI (51), libRoadRunner (36), the Multiscale Object Oriented Simulation Environment (52), and SimBio (53). We have developed an additional package, "mkstd" (github.com/dilpath/mkstd), which we used to implement the EFECT report. This implementation enables the reading, writing, and validation of EFECT reports in any of the HDF5, JSON, XML, and YAML file formats, independently of "mkstd", "libSSR", or Python. Future work should also define the computations and minimal necessary data to recover the probability distribution functions encoded in ECFs according to our standard data, which could provide an efficient means to share statistical model results using an enhanced form of the EFECT Report.

Lastly, future work should promote community adoption of EFECT by simulation tools, public repositories, journals, modeling communities, and regulatory agencies. Simulation tools could provide built-in features that perform our Modeler's workflow. Modeling communities could develop standard test cases like those performed here to develop consensus on EFECT convergence points that define sufficiently reproduced results. Such was how we chose the *ad hoc* convergence point used in this work, which produced marginal false positives (e.g., Figure 1E) in a reasonable amount of time for the modeling methodologies we primarily used. Communities that use expensive or exceptionally stochastic simulations may be inclined to adopt more forgiving criteria for reproducibility. Public repositories could curate stochastic models according to EFECT convergence points and report EFECT errors from testing for reproduced results. Journals could promote or even require including our standard data as supplementary information accompanying publication of stochastic models (3,16). Regulatory agencies could define what EFECT convergence point must be met, or proven, for stochastic simulation results to be considered valid evidence supporting applications and guidance on policymaking.

## Materials and methods

We test for equality in distribution of two stochastic processes $X(t)$ and $Y(t)$ by testing whether the two stochastic processes have the same unknown probability distribution $f(x;\ t)$ (*i.e.*, $X(t) \overset{\mathcal{D}}{=} Y(t)$ if and only if $X, Y \sim f$). Our test is performed over an arbitrary set of simulation times on samples produced from repeated execution of stochastic simulation without regard for determining, or performing any characterization of, the underlying distribution(s). Rather, our method relies on the limit behavior of empirical distributions. Specifically, our method relies characteristic functions (CFs). For a stochastic process $X(t)$, the CF $\phi_X(\tau; t)$ is defined as

$$\phi_X(\tau;t) = \mathbb{E}\left[e^{iX(t)\tau}\right]. \quad [1]$$

Here $\tau$ is a transform variable and $i$ is the imaginary unit. We built our method around CFs because of specific useful properties of a CF: 1. The magnitude of all CFs is in [0,1], which allows assessment of model variables without case-specific decision-making to handle their statistical features (*e.g.*, means at different orders of magnitude); 2. A CF supports all bounded, real-valued random variables, whether discrete or continuous, whereas the probability distribution function only supports continuous random variables; and 3. The empirical characteristic function (ECF) is naturally computed from a sample of stochastic simulation results (54). For a sample $\{X_i(t)\}_{i=1}^{n}$ of size $n$ of $X(t)$ at time $t$, the ECF $\phi_{X,n}(\tau;t)$ is defined as

$$\phi_{X,n}(\tau;t) = \frac{1}{n}\sum_{1\leq j\leq n} e^{iX_j(t)\tau} \quad [2]$$

We use Lévy's Continuity Theorem to test for equality in distribution of simulation results: for samples $\{X_i(t)\}_{i=1}^{n}$ of $X(t)$ and $\{Y_i(t)\}_{i=1}^{n}$ of $Y(t)$, $\{X_i(t)\}_{i=1}^{n} \xrightarrow{\mathcal{D}} X(t)$ if and only if $\lim_{n\to\infty} \phi_{X_n}(t) = \phi_X(t)$, and likewise for $\{Y_i(t)\}_{i=1}^{n}$ and $Y(t)$. For two sampled random processes $X(t)$ and $Y(t)$ over $t \in \mathcal{T}$, we test for pointwise convergence of the characteristic functions $\phi_X(\tau;t)$ and $\phi_Y(\tau;t)$ over $\tau \in \mathcal{t}$ to the same unknown distribution using the EFECT error $\delta$,

$$\delta\left(\phi_{X,n}(\tau;t), \phi_{Y,n}(\tau;t)\right) = \sup\left\{\left|\phi_{X,n}(\tau;t) - \phi_{Y,n}(\tau;t)\right| : \tau \in \mathcal{t}, t \in \mathcal{T}\right\}. \quad [3]$$

It naturally follows that $X(t) \stackrel{\mathcal{D}}{=} Y(t)$ if and only if $\delta\left(\phi_{X,n}(\tau;t), \phi_{Y,n}(\tau;t)\right) \to 0$ as $n \to \infty$. When comparing two samples of a model, we consider the maximum EFECT error of all stochastic variables of the model as the EFECT error of the two samples.

*Methodological Details*

Our test for reproducibility evaluates the likelihood that a sample would be assessed as reproduced by another sample of the same model according to comparison of generated ECFs. A sample of size $2N$ is tested for reproducibility by sampling the EFECT error [ 3 ] via random selection and evaluation of two $N-$ combinations. If the sample produces statistics of the EFECT error that satisfy some *a priori* criterion, an EFECT convergence point, then the sample is considered sufficiently reproducible. While we intend in this work to enable consensus building as to what criterion should be considered broadly acceptable, generally an EFECT error closer to zero produces more reliably reproducible results. Tests thus far suggest that requiring an EFECT error mean below 0.05 may be unnecessarily demanding to detect differences in samples that may not be significant to some modeling communities (Figure 1).

The CF has support for the transform variable (*i.e.*, $\tau$ in [ 1 ] and [ 2 ]) over the real line. Since the intent of the EFECT report is to encode sufficient information to test whether the sample represented by the EFECT report is sufficiently reproduced by another sample, we define a parameterization of the domain of ECFs to produce representative evaluations to test for equality. Specifically, choosing an excessively small domain produces evaluations that are all approximately equal to one, while choosing an excessively large domain can produce evaluations that are equal to either one (*i.e.*, when $\tau = 0$) or zero elsewhere (since, by Riemann-Lebesgue theorem, $\phi_X(\tau;t) \to 0$ as $|\tau| \to \infty$ if $X$ has a density, of which $\phi_X(\tau;t)$ is the Fourier dual). We rewrite [ 1 ] to produce a form for parameterizing the domain over which to evaluate an ECF,

$$\phi_X(\tau;t) = \theta(\tau;t)\sqrt{\Big(\mathbb{E}\big(\cos\big(Z(t)\sigma\big(X(t)\big)\tau\big)\big)\Big)^2 + \Big(\mathbb{E}\big(\sin\big(Z(t)\sigma\big(X(t)\big)\tau\big)\big)\Big)^2}, \qquad [\ 4\ ]$$
$$\sigma\big(X(t)\big)Z(t) = X(t) - \mu\big(X(t)\big).$$

Here $\theta(\tau;t)$ is the phase of $\phi_X(\tau;t)$, the details of which are unimportant here except to note that $|\theta(\tau;t)| = 1$, and $Z(t)$ is standardized $X(t)$ with mean $\mu\big(X(t)\big)$ and standard deviation $\sigma\big(X(t)\big)$. Oscillations in the amplitude of $\phi_X(\tau;t)$ are expected to occur on the scale of periods inversely proportional to the standard deviation of $X(t)$ (since $Z(t)$ is standardized), hence we evaluate an ECF over $m$ periods inversely proportional to the standard deviation,

$$\tau \in \left[0, \frac{2\pi m}{\sigma\big(X(t)\big)}\right], \qquad m \geq 1. \qquad [\ 5\ ]$$

In application, we evaluate ECFs over a uniform distribution of the transform variable according to [ 5 ] for a given value of $m$. All results in this work were performed with a value of $m$ equal to 3 and 100 uniformly distributed values of the transform variable, which were determined *ad hoc* by inspection of results. Generally, greater values of $m$ for the same number of evaluations risks loss of sufficient encoding, and greater numbers of evaluations increases the storage requirements of the EFECT report. We make no formal recommendation here as to what values for these parameters should be broadly acceptable except to note that preliminary tests suggest that a value of $m$ equal to one may be sufficient for a number of evaluations on the order of 100, whereas increasing the number of evaluations nearer to 1,000 may provide marginal improvements to the quality of encoding for at least $m$ equal to three (Source Code S3.28, S3.29).

Testing for reproduced results is performed using the null hypothesis that the EFECT error from comparing Modeler and Curator ECFs was taken from the same distribution as the EFECT error from the Curator's test for reproducibility. Assessment of whether results have been reproduced is performed according to an *a priori* significance level using Chebyshev's inequality for unknown population mean and variance (55). For Curator's and Modeler's stochastic processes $X(t)$ and $Y(t)$, respectively, a Curator performs the test for reproducibility with their model implementation and generates a sample of EFECT error $\delta_{XX}$ with sample size $n$, mean $\bar{\delta}_{XX}$, and variance $\tilde{\delta}_{XX}$. The Curator then generates an EFECT error $\delta_{XY}$ by comparing the ECF(s) of their sample to that(those) reported by the Modeler and calculates the probability that $\delta_{XY}$ was taken from the same distribution as $\delta_{XX}$,

$$\Pr\big(|\delta_{XX} - \bar{\delta}_{XX}| \geq |\delta_{XY} - \bar{\delta}_{XX}|\big) = \frac{1}{n+1}\left\lfloor \frac{n+1}{n}\left(\frac{n^2-1}{n}\frac{\tilde{\delta}_{XX}}{\left(\delta_{XY} - \bar{\delta}_{XX}\right)^2} + 1\right)\right\rfloor. \qquad [\ 6\ ]$$

Hence, the Curator concludes whether they have reproduced the Modeler's results via testing of the null hypothesis for a significance level $\alpha$,

$$X \overset{\mathcal{D}}{=} Y \leftrightarrow \Pr\big(|\delta_{XX} - \bar{\delta}_{XX}| \geq |\delta_{XY} - \bar{\delta}_{XX}|\big) \geq \alpha. \qquad [\ 7\ ]$$

*Implementation Details*

ODEs were executed with libRoadRunner v2.4.0, except those described as being executed in COPASI v4.42. Stochastic ODE simulation with libRoadRunner was performed using the libRoadRunner implementation of the Gillespie stochastic simulation algorithm. ODEs executed in libRoadRunner were specified in Antimony v2.13.4 (56) to generate SBML specification, except those imported from BioModels. ODEs executed in COPASI were specified using generated SBML from Antimony specification. Parameter sampling for ODE simulation with libRoadRunner was performed using methods provided in SciPy v1.11.3 (57). The GSW ABM NetLogo implementation was executed in NetLogo v6.1.1 (September 26, 2019; Source Code S4.2); the MATLAB implementation was executed in MATLAB 2020b (Source Code S4.1); the Python implementation was executed in Python v3.10.13 (Source Code S4.3). All ECF generation and analysis was performed using library functions available in NumPy v1.26.0 (58). The stochastic Boolean network model was executed in the MaBoSS deployment in CompuCell3D v4.5.0 (59). Source code to perform all tests and analyses in Python and simulations with libRoadRunner and MaBoSS are available in Source Code S2. Source code to perform data generation with COPASI is available in Source Code S5. Source code to install all Python dependencies is available in Source Code S1. All project source code for this work is available at https://github.com/tjsego/ssr_project_2024. Instructions to install Python dependencies and execute Python source code and Jupyter Notebooks are available in Instruction S1.

## Conclusion

Reproducibility is essential to scientific credibility, yet stochastic simulations—used across biology, physics, epidemiology, economics, and beyond—have long lacked a general, quantitative standardized framework for assessing reproducibility. EFECT addresses this critical gap by providing a statistically rigorous, model-agnostic method to evaluate whether stochastic simulation results can be independently reproduced. Applicable across modeling paradigms and disciplines, EFECT enables researchers to distinguish meaningful results from artifacts of randomness, thereby improving the reliability of conclusions drawn from stochastic models. By establishing a concrete measure of reproducibility and integrating it into open-source infrastructure and public repositories like BioModels, EFECT lays the foundation for community-driven standards that can elevate the integrity of stochastic studies across all domains of science and engineering.

## Acknowledgments and funding sources

TJS acknowledges funding from National Institutes of Health grant number U24EB028887 and National Science Foundation grant number 2000281. LLF acknowledges funding from National Institutes of Health grants number R01 GM127909 and R01 AI135128, and Defense Advanced Research Projects Agency grant number HR00112220038. ACK acknowledges funding from National Science Foundation grant number 2325776. KT, HH and RMS acknowledge EMBL Core funding. HMS acknowledges funding from National Institutes of Health grant numbers U24EB028887 and P41EB023912. RCL acknowledges funding from National Institutes of Health grants number R01 GM127909, R01 AI135128, and R01 HL169974, and Defense Advanced Research Projects Agency grant number HR00112220038. SR acknowledges funding from the Kavli Foundation under CZI EOSS Cycle-6 grant number LS-2024-GR-34-2950. HEG and MS acknowledge funding from the Universidad de. Buenos Aire and CONICET.

FTB was supported LIBIS, and by de.NBI, the German Network for Bioinformatics Infrastructure (W-de.NBI-016). DP acknowledges funding from the Deutsche Forschungsgemeinschaft (DFG, German Research Foundation) under the project ID 432325352 – SFB 1454, and from the University of Bonn, Germany (via the Schlegel Professorship of Prof. Dr. Jan Hasenauer).

The authors would like to thank Lutz Brusch, Randy Heiland, Jürgen Pahle, Sven Sahle, and Lucian Smith for their insightful feedback during reproducibility workshop discussions at the COMBINE 2023 (University of Connecticut), HARMONY 2024 (University College London), COMBINE 2024 (University of Stuttgart), and HARMONY 2025 (Katholieke Universiteit Leuven) events.

**Competing Interests**

None.

**Supplementary Materials**

- Figure S1. Testing sensitivity of the error metric to number of model variables. A: Oscillator models with increasing number of variables of 2, 3, 4, 5, and 10 were sampled while randomly sampling the frequency of oscillatory variables. B: Box plot of error metrics while sampling for self-similarity for 100, 1k, and 10k sample sizes and 2 (Test 1), 3 (Test 2), 4 (Test 3), 5 (Test 4), and 10 (Test 5) model variables.
- Figure S2. Testing sensitivity of the error metric to differences in orders of magnitude in model variables. A: A biphasic variable (left) controls the frequency of an oscillatory variable (right) with amplitude 1 (Test 1), 1E-3 (Test 2), or 1E-6 (Test 3). B: Box plot of error metrics while sampling for self-similarity for 100, 1k, and 10k sample sizes and oscillatory variable amplitude 1 (Test 1), 1E-3 (Test 2), or 1E-6 (Test 3).
- Figure S3. Process diagram of reproduced stochastic simulation results using EFECT. Modeler Workflow: A modeler (i.e., the producer of stochastic results) iteratively samples a stochastic model, or stochastic model inputs, until the aggregated sample is sufficiently reproducible. After showing reproducibility of results, the modeler reports minimal supporting data along with their publication that facilitates reproducing results by others. Curator Workflow: A curator (i.e., the reproducer of stochastic results) reproduces published results by implementing the published model, generating a sample, and testing for equality in distribution of their results according to the supporting data that accompanies the model.
- Instruction S1. Instructions for software installation and execution.
- Source Code S1. setup/: Source code to install all software.
- Source Code S2. code/: Source code implementing tests, simulations, tested models, and method equations.
    - Source Code S2.1. cc3d_cell_fate.py: Data generation for testing the approach on results from a stochastic Boolean network model.
    - Source Code S2.2. perf_compare_size_1.py: Runtime test for computational cost of the approach with increasing number of model variables.
    - Source Code S2.3. sim_2_modeler_py.py: Simulation of the Modeler’s workflow. Generates data for use in simulation of the Curator’s workflow for cases where data with six and nine significant figures was used.
    - Source Code S2.4. sim_2_curator_py.py: Simulation of the Curator’s workflow. Uses generated data from simulation of the Modeler’s workflow and externally generated data from COPASI. Tests cases where results are successfully reproduced, unsuccessfully

reproduced due to differences in parameter distributions, and unsuccessfully reproduced due to significant figures of data from different simulators.
    - Source Code S2.5. test_comparison.py: Tests the likelihood of detecting parameter differences in a viral infection model as a function of sample size and scale of difference in distribution mean and standard deviation.
- Source Code S3. notebooks/: Jupyter Notebooks demonstrating various test cases and simulation of the reproducible simulation workflow.
    - Source Code S3.1. proto_paramvar.ipynb: Demonstration of basic approach with a viral/bacterial coinfection model while sampling one parameter with a normal distribution.
    - Source Code S3.2. proto_2paramvar.ipynb: Demonstration of basic approach with a viral/bacterial coinfection model while sampling two parameters with a normal distribution.
    - Source Code S3.3. proto_10paramvar.ipynb: Demonstration of basic approach with a viral/bacterial coinfection model while sampling ten parameters with a normal distribution.
    - Source Code S3.4. proto_10paramvar2.ipynb: Demonstration of basic approach with a viral/bacterial coinfection model while sampling ten parameters with a uniform distribution.
    - Source Code S3.5. proto_bistable.ipynb: Demonstration of basic approach with a two-variable biphasic/oscillatory model while sampling one initial condition with a uniform distribution.
    - Source Code S3.6. proto_bistable2.ipynb: Demonstration of basic approach with a two-variable biphasic/oscillatory model while sampling one parameter with a normal distribution.
    - Source Code S3.7. proto_liu2012.ipynb: Demonstration of basic approach with BioModels entry MODEL2004140002.
    - Source Code S3.8. proto_lo2005.ipynb: Demonstration of basic approach with BioModels entry MODEL1805160001.
    - Source Code S3.9. proto_lorenz.ipynb: Demonstration of basic approach with a Lorenz system model using a stochastic simulation algorithm.
    - Source Code S3.10. proto_pendulum.ipynb: Demonstration of basic approach with a nonlinear pendulum model with two cases of sampling an initial condition with a uniform distribution, and one case of sampling a parameter with a normal distribution.
    - Source Code S3.11. proto_pulse.ipynb: Demonstration of basic approach with a pulse-like oscillatory model while sampling one parameter with a normal distribution.
    - Source Code S3.12. proto_seir.ipynb: Demonstration of basic approach with a viral infection model using a stochastic simulation algorithm.
    - Source Code S3.13. proto_compare_1.ipynb: Demonstration of detecting structural differences in models, one of which is subjected to parameter sampling, and the other of which is simulated with a stochastic simulation algorithm.
    - Source Code S3.14. proto_compare_2.ipynb: Demonstration of detecting parametric differences for a viral infection model subject to parameter sampling. Parametric differences are implemented as proportionally scaled differences in parameter distribution mean and standard deviation.
    - Source Code S3.15. proto_compare_3.ipynb: Demonstration of detecting parametric differences in constant-valued models. Parametric differences are implemented as differences in line intercept.

- Source Code S3.16. proto_compare_4.ipynb: Demonstration of detecting parametric differences for a two-variable biphasic oscillatory model subject to parameter sampling. Parametric differences are implemented as differences in parameter distribution mean.
- Source Code S3.17. proto_compare_5.ipynb: Demonstration of detecting parametric differences for a reaction-kinetics model subject to parameter sampling. Parametric differences are implemented as proportionally scaled differences in parameter distribution mean and standard deviation.
- Source Code S3.18. proto_compare_5a.ipynb: Demonstration of detecting parametric differences for a reaction-kinetics model simulated with a stochastic simulation algorithm.
- Source Code S3.19. proto_compare_6.ipynb: Demonstration of detecting parametric differences for a viral infection model subject to parameter sampling. Parametric differences are implemented as differences in parameter distribution standard deviation.
- Source Code S3.20. proto_compare_7.ipynb: Demonstration of summary statistics incorrectly assessing samples with different initial conditions as equal.
- Source Code S3.21. gsw.ipynb: Demonstration of basic approach with three implementations of a GSW ABM.
- Source Code S3.22. sim_modeler_2.ipynb: Simulation of the Modeler's workflow when generating reproducible simulation results.
- Source Code S3.23. sim_curator_2_pass.ipynb: Simulation of the Curator's workflow when successfully reproducing simulation results using supporting data generated from the Modeler's workflow.
- Source Code S3.24. sim_curator_2_fail_params.ipynb: Simulation of the Curator's workflow when unsuccessfully reproducing simulation results using supporting data generated from the Modeler's workflow. Results are not reproduced due to differences in a model parameter distribution.
- Source Code S3.25. sim_curator_2_fail_sigfig.ipynb: Simulation of the Curator's workflow when unsuccessfully reproducing simulation results using supporting data generated from the Modeler's workflow. Results are not reproduced due to differences in significant figures of data output by the simulators used in the Modeler's and Curator's workflows.
- Source Code S3.26. proto_compare_boolean.ipynb: Demonstration of basic approach with a stochastic Boolean network model.
- Source Code S3.27. proto_compare_mag_1.ipynb: Test for sensitivity of the method to differences in magnitude of model variables.
- Source Code S3.28. proto_compare_periods_1.ipynb: Test for sensitivity of the method to different periods of evaluation of ECFs.
- Source Code S3.29. proto_compare_res_1.ipynb: Test for sensitivity of the method to different discretization resolutions of the transform variable when evaluating ECFs.
- Source Code S3.30. proto_compare_size_1.ipynb: Test for sensitivity of the method to different numbers of model variables.
- Source Code S3.31. proto_compare_var_2.ipynb: Test for sensitivity of the method to different orders of magnitude of the standard deviation of a parameter distribution.
- Source Code S3.32. pandemic_adak2020.ipynb: Assessment of the reproducibility of results from a pandemic stochastic differential equation model.
- Source Code S3.33. pandemic_din2020.ipynb: Assessment of the reproducibility of results from a pandemic stochastic differential equation model.

- Source Code S3.34. pandemic_faranda2020.ipynb: Assessment of the reproducibility of results from a pandemic stochastic differential equation model.
  - Source Code S3.35. pandemic_mamis2022.ipynb: Assessment of the reproducibility of results from a pandemic stochastic differential equation model.
  - Source Code S3.36. pandemic_ninotorres2022.ipynb: Assessment of the reproducibility of results from a pandemic stochastic differential equation model.
  - Source Code S3.37. pandemic_tesfaye2021.ipynb: Assessment of the reproducibility of results from a pandemic stochastic differential equation model.
  - Source Code S3.38. finance_liesenfeld2003.ipynb: Assessment of the reproducibility of results from a finance stochastic differential equation model.
  - Source Code S3.39. spde_zhang2019.ipynb: Assessment of the reproducibility of results from a stochastic partial differential equation model.
  - Source Code S3.40. compare_moose2.ipynb: Comparison of stochastic results produced by libRoadRunner and the Multiscale Object Oriented Simulation Environment.
  - Source Code S3.41. compare_simbio_roadrunner.ipynb: Generation and comparison of stochastic results with libRoadRunner and SimBio.
  - Source Code S3.42. simulators.ipynb: Example reproducibility workflow with the COPASI Python API (BasiCO) and libRoadRunner using the tools and interface provided by the Python implementation of EFECT in libSSR.
- Source Code S4. gsw/: Source code for GSW ABM implementations.
  - Source Code S4.1. GSW_MatLab_source: MATLAB implementation of the GSW ABM.
  - Source Code S4.2. GSW_NetLogo_source: NetLogo implementation of the GSW ABM.
  - Source Code S4.3. GSW_Python_source: Python implementation of the GSW ABM.
- Source Code S5. copasi/: Source code generating simulation data for the Curator's workflow.
  - Source Code S5.1. curator_results_diff/: Source code to generate results with COPASI when simulating the Curator's workflow with no model differences.
  - Source Code S5.2. curator_results_same/: Source code to generate results with COPASI when simulating the Curator's workflow with parametric differences.
- Source Code S6. compare_moose2/: Results produced from the Multiscale Object Oriented Simulation Environment for comparison in Source Code S3.40.
- Table S1. Summary of all tests and simulation experiments.
- Table S2. Details on all tests and simulation experiments, including tested sample size(s), modeling methodology, and name of source code in Supplementary Materials.
- Table S3. Summary and examples of the minimal data set supporting information exchange for reproducible stochastic simulation results.